\documentclass[aps,showpacs,superscriptaddress,amsfonts,amssymb,amsmath]{revtex4}

\begin{document}

\preprint{Yukawa Institute Kyoto} \preprint{YITP-07-10}

\title{Deformed multi-variable Fokker-Planck equations}

\author{Choon-Lin Ho}
 \affiliation{Department of Physics, Tamkang University,
 Tamsui 251, Taiwan, Republic of China}
\author{Ryu  Sasaki}
 \affiliation{Yukawa Institute for Theoretical Physics,
     Kyoto University, Kyoto 606-8502, Japan}

\date{Mar 9, 2007}

\begin{abstract}

In this paper new multi-variable deformed Fokker-Planck (FP)
equations are presented.  These deformed FP equations are
associated with the Ruijsenaars-Schneider-van Diejen (RSvD) type
systems in the same way that the usual one variable FP equation is
associated with the one particle Schr\"odinger equation.  As the
RSvD systems are the ``discrete" counterparts of the celebrated
exactly solvable many-body Calogero-Sutherland-Moser systems, the
deformed FP equations presented here can be considered as
``discrete" deformations of the ordinary multi-variable FP
equations.

\end{abstract}


 \maketitle

\section{Introduction} The Fokker-Planck (FP) equation is
 one of the most important tools to deal with
fluctuation phenomena in various kinds of systems \cite{FP}.
Recent discovery of anomalous diffusion in fractal and disordered
media have prompted new developments in the FP theory, the major
one being fractional FP equations, where the ordinary spatial
derivatives are replaced by fractional derivatives \cite{FracFP}.
Nonlinear FP equations \cite{NFP} and diffusion equation based on
$q$-derivatives have also been considered \cite{Ho}.

In \cite{HS}, based on the well-known relation {between FP
equations and Schr\"odinger equations of quantum mechanics, we
have proposed new types of deformed FP equations which are
associated with the Schr\"odinger equations of the ``discrete"
quantum mechanics considered in \cite{OS1,OS2,OS3}. The latter is
a natural discretization of quantum mechanics and its
Schr\"odinger equations are difference instead of differential
equations. The eigenfunctions of some exactly solvable ``discrete"
quantum mechanics include various deformations of the classical
orthogonal polynomials (the Hermite, Laguerre and Jacobi
polynomials), namely, those belonging to the family of
Askey-scheme of hypergeometric orthogonal polynomials, and the
Askey-Wilson polynomial in $q$-analysis \cite{AAR,KS,AW}.  In
fact, these deformed orthogonal polynomials arise in the problem
\cite{OS2} of describing the equilibrium positions of
Ruijsenaars-Schneider-van Diejen (RSvD) type systems \cite{RSD},
which are the ``discrete" counterparts of the celebrated exactly
solvable multi-particle Calogero-Sutherland-Moser (CSM) systems
\cite{CSM}, just as the Hermite, Laguerre and Jacobi polynomials
describe the equilibrium positions of the CSM systems \cite{zero}.
In this sense, the proposed deformed FP equations \cite{HS} can be
considered as the ``discrete" deformation of the usual FP
equation.

In this paper, we would like to generalize the results in
\cite{HS} for the one-particle cases to the many-body cases, and
obtain the many-variable deformed FP equations associated with the
RSvD systems.  We first briefly review in Sect. II the connection
between the ordinary multi-variable FP equation and the many-body
Schr\"odinger equation.  Sect. III summarizes the basics of the
CSM systems associated with rational/trigonometric classical root
systems. In Sect. IV we review the RSvD systems and derive the
corresponding FP equations for the two rational types. The
trigonometric type RSvD systems and the corresponding FP equations
are discussed in Sect. V.  Sect. VI concludes the paper.

\medskip

\section{Connection between Fokker-Planck and Schr\"odinger equations}

The general form of FP equation of the probability density
$P(x,t)$ with $n$ variables $x\equiv {}^t(x_1,x_2,\ldots,x_n)$ is
\cite{FP}
\begin{gather}
\frac{\partial P(x,t)}{\partial t}=L_{FP}P(x,t),\nonumber\\
L_{FP}\equiv -\sum_j\frac{\partial}{\partial x_j} D_j(x) +
\sum_{jk}\frac{\partial^2}{\partial x_j\partial x_k}D_{jk}(x).
 \label{FPE}
\end{gather}
The functions $D_j(x)$ and $D_{jk} (x)$ in the FP operator
$L_{FP}$ are, respectively,  the drift vector and the diffusion
matrix (we consider only time-independent case).  The drift vector
represents the external forces acting on the particles, while the
diffusion matrix accounts for the effects of fluctuation.

It is well known that in one-dimension, the FP equation is closely
related to the Schr\"odinger equation. The correspondence between
the two equations can be shown by transforming the FP equation
into the corresponding Schr\"odinger equation \cite{FP}, or vice
versa \cite{HS}, based on a similarity transformation. The
similarity transformation is related to the ground state of the
Schr\"odinger equation, or equivalently, to the stationary state
of the FP equation. One thus expects that similar connection is
possible between a FP equation with multiple variables and a
one-dimensional many-body Schr\"odinger equation.  It turns out
that such connection is possible only for special forms of $D_j$
and $D_{jk}$ \cite{FP}.  Below we shall briefly review such
connection.

Instead of transforming an assumed form of FP equation to a
Schr\"odinger equation, as is usually done in the literature (see
e.g. \cite{FP}), we think it is more direct to see what form of FP
equation will emerge by directly transforming a known many-body
Schr\"odinger equation.  So let us consider a quantum mechanical
Hamiltonian of $n$ particles with equal masses (for simplicity, we
adopt the unit system in which $\hbar$, the mass $m$ of the
particles, and  the `fictitious' speed of light $c$ in the RSvD
systems are such that $\hbar=2m=mc=1$)
\begin{equation}
H=-\sum_{i=1}^n \frac{\partial^2}{\partial x_i^2} + V(x).
\end{equation}
Here $V(x)$ is the potential term which include external fields
and inter-particle forces. By adjusting the additive constant in
$V(x)$,  the energy of the ground state $\phi_0$ can always be
made zero, i.e. $H\phi_0=0$. Under these assumptions, the
Hamiltonian $H$ is completely determined by its ground state wave
function $\phi_0(x)$. By the well-known theorem of quantum
mechanics, $\phi_0(x)$ has no node and can be chosen real. Hence
one can parametrize $\phi_0$ by a real prepotential $W(x)$:
\begin{equation}
\phi_0(x)=e^{-W(x)}. \label{W}
\end{equation}
 Then the condition
$H\phi_0=0$ enables one to express $V(x)$ in terms of $W(x)$ as
well:
\begin{equation}
V(x)=\sum_{j=1}^n \biggl(
  \Bigl(\frac{\partial W(x)}{\partial x_j}\Bigr)^2
  -\frac{\partial^2 W(x)}{\partial x_j^2}\biggr).
\label{V}
\end{equation}
Since only derivatives of $W(x)$ appear in $V(x)$, $W(x)$ is
defined only up to an additive constant $W_0$.  We choose the
constant $W_0$ in such a way as to normalize $\phi_0(x)$ properly,
$\int\phi_0(x)^2\,d^nx=1$. For simplicity of presentation, we
consider the cases in which the ground state wave functions are
square integrable, that is the corresponding FP operators have the
normalizable stationary distributions.

The Hamiltonian $H$ can be expressed as a sum of factorized forms,
$H= \sum_j A_j^\dagger A_j$, where
\begin{gather}
A_j\equiv  \frac{\partial}{\partial x_j} + \frac{\partial
W(x)}{\partial x_j},\label{A}\\
 A_j^\dagger\equiv -\frac{\partial}{\partial x_j}
 +\frac{\partial W(x)}{\partial x_j}.\nonumber
 \end{gather}
Since each term $A_j^\dagger A_j$ in $H$ is positive
semi-definite, the condition $A_j\phi_0=0$ must hold in order to
have  $H\phi_0=0$. This latter condition is a sufficient and
necessary condition for $H\phi_0=0$.

Now we define an operator from the $H$ and $\phi_0$ by the
similarity transformation
\begin{equation}
L_{FP}\equiv -\phi_0 H \phi_0^{-1},\label{FP}
\end{equation}
which guarantees the non-positivity of the eigenvalues of
$L_{FP}$. Using $\phi_0$ in (\ref{W}), one obtains
\begin{equation}
L_{FP}=\sum_j\frac{\partial}{\partial
x_j}\left(\frac{\partial(2W(x))}{\partial x_j}\right) + \sum_j
\frac{\partial^2}{\partial x_j^2}\,.
\end{equation}
{}From (\ref{FPE}), it is seen that $L_{FP}$ is just the
corresponding FP operator with drift vector $D_j=-2{\partial
W}/{\partial x_j}$ and a diffusion matrix proportional to the unit
matrix: $D_{jk}=\delta_{jk}$ (the constant factor here equals
unity which is a result of the choice of unit in $H$).  This
establishes the required forms of $D_j$ and $D_{jk}$ so that the
multi-variable FP equation is related to a one-dimensional
many-body Schr\"odinger equation.  We see that $D_j$ is in fact
defined, as in the one variable FP equation, by a drift potential
$\Phi (x)=2W(x)$: $D_j=-{\partial \Phi}/{\partial x_j}$.

Both $H$ and $L_{FP}$ are determined by $\phi_0$. The relationship
between their eigenfunctions is the same as that for the single
particle case, except that special care is needed to ensure the
reality of the eigenfunctions belonging to degenerate eigenvalues.
The eigenfunction $P_{\bf m} (x)$ of $L_{FP}$ corresponding to the
eigenvalue $-\lambda_{\bf m}$ is related to the (real and
normalized) eigenfunction $\phi_{\bf m}$ of $H$ corresponding to
$\lambda_{\bf m}$ by $P_{\bf m}(x)=\phi_0(x)\phi_{\bf m}(x)$. Here
${\bf m}$ is, in general, a certain multi-index. The stationary
distribution is $P_0=\phi_0^2=\exp(-2W)$, which is obviously
non-negative, and is the zero mode of $L_{FP}$: $L_{FP}P_0=0$. Any
positive definite initial probability density $P(x,0)$ can be
expanded as $P(x,0)=\phi_0(x)\sum_{\bf m} c_{\bf m}\phi_{\bf
m}(x)$, with constant coefficients
$c_{\bf m}$ 
\begin{eqnarray}
c_{\bf m}=\int_{-\infty}^\infty \phi_{\bf
m}(x)\left(\phi_0^{-1}(x) P(x,0)\right)d^nx.
\end{eqnarray}
Then at any later time $t$, the solution of the FP equation is
$P(x,t)=\phi_0(x)\sum_{\bf m} c_{\bf m} \phi_{\bf
m}(x)\exp(-\lambda_{\bf m} t)$.

\section{Calogero-Sutherland-Moser systems}

One class of multi-variable FP equations having the special form
discussed in the previous section which has been widely discussed
in the literature is that related to the celebrated exactly
solvable many-body quantum system, namely, the CSM models. Such FP
equation has found applications in, e.g. random matrix theory
\cite{Dyson} and others \cite{n-s}.

Here we consider four  types  of CSM systems associated with the
classical root systems: two rational and two trigonometric types.
As mentioned earlier, we consider only the cases with explicitly
known ground state wave functions $\phi_0(x)$, which are square
integrable. This excludes the other well-known cases of the
hyperbolic and elliptic potentials as well as the rational cases
without the harmonic confining potential. The potential $V_{\rm
CS}(x)$ of the CSM systems can be written in terms of the
prepotential $W(x)$ as in (\ref{V}). The explicit forms of the
four types of potentials $V_{\rm CS}(x)$ and the corresponding
prepotentials $W(x)$ are as follows:

\vskip 0.8cm \noindent (\romannumeral1) rational $A_{n-1}$ :
\begin{subequations}
\begin{align}
  V_{\rm CS}(x)&=\sum_{j=1}^n x_j^2
  +\sum_{\genfrac{}{}{0pt}{2}{j,k=1}{j\neq k}}^n
  \frac{g(g-1)}{(x_j-x_k)^2}
  - n\bigl(1+g(n-1)\bigr),
  \label{ratA_CS_V}\\
  W(x)&=\frac{1}{2}\sum_{j=1}^n x_j^2 - g\sum_{1\leq j<k\leq n}
  \log\bigl|x_j-x_k\bigr|+W_0,
  \label{ratA_CS_W}
\end{align}
\end{subequations}
(\romannumeral2) rational $BC_n$ :
\begin{subequations}
\begin{align}
  V_{\rm CS}(x)&=\sum_{j=1}^n\biggl(x_j^2
   + \frac{g_S(g_S-1)}{x_j^2}\biggr)
  + g_M(g_M-1)\sum_{\genfrac{}{}{0pt}{2}{j,k=1}{j\neq k}}^n
  \biggl(\frac{1}{(x_j-x_k)^2}+\frac{1}{(x_j+x_k)^2}
  \biggr)\nonumber\\
  &\quad
  -2n\bigl(g_S+\tfrac{1}{2}+g_M(n-1)\bigr),
  \label{ratBC_CS_V}\\
  W(x)&=\frac{1}{2}\sum_{j=1}^n x_j^2-g_M\sum_{1\leq j<k\leq n}
   \Bigl(\log \bigl|x_j-x_k\bigr|
  +\log \bigl|x_j+x_k\bigr|\Bigr)
%
- g_S \sum_{j=1}^n
  \log \bigl|x_j\bigr|+W_0,
  \label{ratBC_CS_W}
\end{align}
\end{subequations}
(\romannumeral3) trigonometric $A_{n-1}$ :
\begin{subequations}
\begin{align}
  V_{\rm CS}(x)&=\frac{\pi^2}{L^2}
g(g-1)\sum_{\genfrac{}{}{0pt}{2}{j,k=1}{j\neq k}}^n
  \frac{1}{\sin^2\frac{\pi}{L}(x_j-x_k)}-\frac{\pi^2 g^2}{L^2}
  \frac{n(n^2-1)}{3},
  \label{trigA_CS_V}\\
  W(x)&=-g\sum_{1\leq j<k\leq n}
  \log\bigl|\sin\tfrac{\pi}{L}(x_j-x_k)\bigr|+W_0,
  \label{trigA_CS_W}
\end{align}
\end{subequations}
(\romannumeral4) trigonometric $BC_n$ :
\begin{subequations}
\begin{align}
  V_{\rm CS}(x)&=\frac{ \pi^2}{L^2}\sum_{j=1}^n\biggl(
  \frac{(g_S+g_L)(g_S+g_L-1)}{\sin^2\frac{\pi}{L}x_j}
  +\frac{g_L(g_L-1)}{\cos^2\frac{\pi}{L}x_j}\biggl)\nonumber\\
  &\quad
  +\frac{\pi^2}{L^2}\sum_{\genfrac{}{}{0pt}{2}{j,k=1}{j\neq k}}^n
  \biggl(\frac{g_M(g_M-1)}{\sin^2\frac{\pi}{L}(x_j-x_k)}
  +\frac{g_M(g_M-1)}{\sin^2\frac{\pi}{L}(x_j+x_k)}\biggr)\nonumber\\
  &\quad
  -\frac{\pi^2 n}{L^2}\Bigl(\bigl(g_S+2g_L+g_M(n-1)\bigr)^2
  +\frac{g_M^2}{3}(n^2-1)\Bigr),
  \label{trigBC_CS_V}\\
  W(x)&=-\sum_{1\leq j<k\leq n}g_M\Bigl(
  \log\bigl|\sin\tfrac{\pi}{L}(x_j-x_k)\bigr|
  +\log\bigl|\sin\tfrac{\pi}{L}(x_j+x_k)\bigr|\Bigr)\nonumber\\
  &\quad
  -\sum_{j=1}^n\Bigl(g_S\log\bigl|\sin\tfrac{\pi}{L}x_j\bigr|
  +g_L\log\bigl|\sin\tfrac{\pi}{L}2x_j\bigr|\Bigr)+W_0.
  \label{trigBC_CS_W}
\end{align}
\end{subequations}

Here $W_0$ is the constant term necessary for the normalization of
the ground state wave function $\phi_0(x)$. The constant terms in
$V_{\rm CS}(x)$ are the consequences of the expression (\ref{V})
in terms of the prepotential. In the above formulas $g,g_S,g_M$
and $g_L$ are non-negative coupling constants, and $L$ is the
circumference of the circle in which $x_j$ live for the
trigonometric cases. In the rational potential cases, (i) and
(ii), the coefficient of the harmonic confining potential $x_j^2$
is usually written as $\frac{1}{2}m\omega^2$. Here we choose
$2m=m\omega=1$ for simplicity.
The rational $D_n$ model is obtained from the rational $BC_n$
model by putting $g_S=0$ in \eqref{ratBC_CS_V} and
\eqref{ratBC_CS_W}. The rational $B_n$, $C_n$ and $BC_n$ models
are all equivalent. The trigonometric $D_n$ model is obtained from
the trigonometric $BC_n$ model by putting $g_S=g_L=0$ in
\eqref{trigBC_CS_V} and \eqref{trigBC_CS_W}, whereas the
trigonometric $B_n$ and $C_n$ models are obtained by $g_L=0$ and
$g_S=0$, respectively.

If we apply the transformation (\ref{FP}) to the Hamiltonian of
the rational $A_{n-1}$ CSM system, the resulted FP equation is
\begin{equation}
L_{FP}=\sum_{j=1}^n \frac{\partial^2}{\partial x_j^2} +
2\sum_{j=1}^n\frac{\partial}{\partial x_j}\left(x_j
-g\sum_{\genfrac{}{}{0pt}{2}{k=1}{j\neq k}}^n
  \frac{1}{x_j-x_k}\right).
  \label{FP-CMS}
\end{equation}
Multi-variable FP equation of this type was first employed by
Dyson as a Brownian motion model of certain random matrix ensemble
\cite{Dyson}.  FP equations corresponding to the other three types
of CSM systems can be obtained accordingly.

\section{Ruijsenaars-Schneider-van Diejen systems}

Now we would like to derive new types of multi-variable FP
equations corresponding to certain discrete deformations of the
above CSM systems; namely, the Ruijsenaars-Schneider-van Diejen
(RSvD) systems with the rational/trigonometric potentials.

The RSvD system is an integrable deformation of the CSM system.
The Hamiltonian has the general form
\begin{eqnarray}
  H =\sum_{j=1}^n \left(\sqrt{V_j(x)}\,e^{-i\partial_j}\sqrt{V_j^*(x)}
  +  \sqrt{V_j^*(x)}\,e^{i\partial_j}\sqrt{V_j(x)}-V_j(x)-V_j^*(x)\right).
  \label{H_RS}
\end{eqnarray}
Here $\partial_j\equiv \partial/\partial{x_j}$, and $V_j(x)$ is
given by
\begin{equation}
  V_j(x)=w(x_j)\prod_{\genfrac{}{}{0pt}{2}{k=1}{k\neq j}}^n
  v(x_j-x_k)\times
  \begin{cases}
  1&\text{for $A_{n-1}$,}\\v(x_j+x_k)&\text{for $BC_n$}.
  \label{V_RS}
  \end{cases}
\end{equation}
We use the conventional notation that $V_j^*(x)$ is the complex
conjugate function of $V_j(x)$. (For an arbitrary function
$f(x)=\sum_na_n x^n$, $a_n\in\mathbb{C}$ we define
$f^*(x)=\sum_na_n^*x^n$. Here $c^*$ is the complex conjugation of
a number $c\in\mathbb{C}$.) Since the operators $e^{\pm
i\partial_j}$ cause finite shifts of the wave function in the
imaginary direction ($e^{\pm i\partial_j}f(x)= f(x_1,\cdots,x_j\pm
i,\cdots,x_n)$), these systems can thus be called ``discrete"
dynamical systems.

The basic potential functions $v(x)$ and $w(x)$ are as follows:

\vskip 0.8cm \noindent (\romannumeral1) rational $A_{n-1}$ :
\begin{subequations}
\begin{align}
  v(x)&=1-i\frac{g}{x},
  \label{ratA_RS_v}\\
  w(x)&=\Bigl(a_1+ix\Bigr)
  \Bigl(a_2+ix\Bigr),
  \label{ratA_RS_w}
\end{align}
\end{subequations}
(\romannumeral2) rational $BC_n$ :
\begin{subequations}
\begin{align}
  v(x)&=1-i\frac{g}{x},
  \label{ratBC_RS_v}\\
  w(x)&=\frac{(a_1+ix)(a_2+ix)(a_3+ix)(a_4+ix)}{2ix(2ix+1)},
  \label{ratBC_RS_w}
\end{align}
\end{subequations}
(\romannumeral3) trigonometric $A_{n-1}$ :
\begin{subequations}
\begin{align}
 \hspace*{-4mm} v(x)&=\frac{\sin\frac{\pi}{L}(x-ig_0)}{\sin\frac{\pi}{L}x},
  \qquad \qquad w(x)=1,
  \label{trigA_RS_vw}\\
%
\end{align}
\end{subequations}
(\romannumeral4) trigonometric $BC_n$ :
\begin{subequations}
\begin{align}
  v(x)&=\frac{\sin\frac{\pi}{L}(x-ig_0)}{\sin\frac{\pi}{L}x},
  \label{trigBC_RS_v}\\
  w(x)&=
  \frac{\sin\frac{\pi}{L}(x-ig_1)}{\sin\frac{\pi}{L}x}\,
  \frac{\sin\frac{\pi}{L}(x-\frac{i}{2}-ig_2)}
       {\sin\frac{\pi}{L}(x-\frac{i}{2})}\nonumber\\
  &\quad\times
  \frac{\cos\frac{\pi}{L}(x-ig'_1)}{\cos\frac{\pi}{L}x}\,
  \frac{\cos\frac{\pi}{L}(x-\frac{i}{2}-ig'_2)}
       {\cos\frac{\pi}{L}(x-\frac{i}{2})}.
  \label{trigBC_RS_w}
\end{align}
\end{subequations}
Here the coupling constants $a_1, a_2,a_3,a_4,g,g_0,g_1,g_2,g'_1$
and $g'_2$ are assumed to be non-negative and one of
$a_1,\ldots,a_4$ must be greater than $1/2$. Apparently, the
deformed theories have usually more coupling constants than the
original ones.

Note that the Hamiltonian \eqref{H_RS} can be expressed as a sum
of factorized forms, $H=\sum_j A_j^\dagger A_j$, with
\begin{gather}
 A_j\equiv e^{-\frac{i}{2}\partial_j}\sqrt{V_j^*(x)}
  -e^{\frac{i}{2}\partial_j}\sqrt{V_j(x)},\\
  A_j^{\dagger}\equiv \sqrt{V_j(x)}\,e^{-\frac{i}{2}\partial_j}
  -\sqrt{V_j^*(x)}\,e^{\frac{i}{2}\partial_j}.
\end{gather}
It should be  remarked that in the small momentum (the large $c$)
limit, in which the shift operators $e^{\pm i\partial_j}$ can be
approximated by differential operators $e^{\pm i\partial_j}\approx
1\pm i
\partial_j-\frac{1}{2}\partial_j^2$, the RSvD systems reduce to
the CSM systems \cite{OS2,OS3}.

We want now to derive new multi-variable FP equations associated
with these RSvD systems, following the procedure described before.
Let us note here that while the trigonometric cases can be treated
exactly in the manner to be discussed below, it is more suitable
to be treated as  multiplicative shift systems instead of additive
ones like the other two cases.  For this reason, we will discuss
the two rational cases first.

The ground state eigenfunctions $\phi_0(x)$ of \eqref{H_RS} for
the rational cases are
\begin{subequations}
\begin{align}
  \text{(\romannumeral1)}\,:\,\ &
  \phi_0(x)\propto \biggl|\,
  \prod_{j=1}^n\Gamma(a_1+ix_j)\Gamma(a_2+ix_j)\,\cdot\!\!\!\!
  \prod_{1\leq j<k\leq n}\frac{\Gamma(g+i(x_j-x_k))}{\Gamma(i(x_j-x_k))}\,
  \biggr|,\label{wf1}\\
  \text{(\romannumeral2)}\,:\,\ &
  \phi_0(x)\propto \biggl|\, \prod_{j=1}^n
  \frac{\prod_{\alpha=1}^4\Gamma(a_{\alpha}+ix_j)}
  {\Gamma(2ix_j)}\,\cdot\!\!\!\!
  \prod_{1\leq j<k\leq n}\prod_{\epsilon=\pm 1}
  \frac{\Gamma(g+i(x_j+\epsilon\,x_k))}{\Gamma(i(x_j+\epsilon\,x_k))}\,
  \biggr|\label{wf2},
\end{align}
\end{subequations}
where $|f(z)|=\sqrt{f(z)^*f(z)}$ for any complex function $f(z)$.
These ground states have zero energy: $H\phi_0=0$. Since $H$
consists of a sum of positive semi-definite terms $A_j^\dagger
A_j$, the condition of zero ground state energy implies
$A_j\phi_0=0$ ($j=1,\ldots,n$), or explicitly,
\begin{gather}
  \!\sqrt{V_j^*\left(x_1,\cdots,x_j-\frac{i}{2},\cdots,x_n\right)}~
  \phi_0\left(x_1,\cdots,x_j-\frac{i}{2},\cdots,x_n\right)\nonumber\\
  \!=\!\sqrt{V_j\left(x_1,\cdots,x_j+\frac{i}{2},\cdots,x_n\right)}~
  \phi_0\left(x_1,\cdots,x_j+\frac{i}{2},\cdots,x_n\right).
\label{Cond}
\end{gather}
This can be verified explicitly using the forms of $V(x)$ and
$\phi_0$ for the two cases.

Now, we form the associated FP operator from (\ref{H_RS}) and
$\phi_0$ in (\ref{wf1}) and (\ref{wf2}) according to the
similarity transformation (\ref{FP}).  With the help of
(\ref{Cond}), we find that
\begin{gather}
  L_{FP}=-\sum_{j=1}^n\left(e^{i\partial_j}V_j(x) +
  e^{-i\partial_j}V_j^*(x)
  - V_j(x)-V_j^*(x)\right).
  \label{FP1}
\end{gather}
This is the general form of FP operator corresponding to the
discrete Hamiltonian $H$ in (\ref{H_RS}). One has
$L_{FP}\phi_0^2=0$ as a consequence of $H\phi_0=0$.  Thus
$\phi_0^2$ is the stationary solution of the respective FP
equation.

In the small momentum limit, or equivalently, the large $c$ limit,
all $x$ and other parameters in $V(x)$ are considered small
quantities \cite{OS2,OS3}.  Eq.~(\ref{FP1}) then reduces to the
corresponding FP equation associated with the CSM system.
Explicitly, keeping in the operators $\exp(\pm i\partial_j)$ only
up to the second order terms in the momentum, we get the FP
operator
\begin{eqnarray}
L_{FP}=\sum_{j=1}^n \left(\frac{\partial^2}{\partial x_j^2}
Re~V_j(x) + \frac{\partial}{\partial x_j}
2Im~V_j(x)\right).\label{n-rel}
\end{eqnarray}
Here in $V_j$ only the lowest order terms in $x$ and other
parameters shall be retained.

As an illustration, let us take the case of rational $A_{n-1}$
type. The potential is
\begin{equation}
V_j(x)=\left(a_1 + ix_j\right)\left(a_2 +
ix_j\right)\prod_{\genfrac{}{}{0pt}{2}{k=1}{k\neq
j}}^n\left(1-i\frac{g}{x_j-x_k}\right).
\end{equation}
In the above-mentioned limit, we have
\begin{eqnarray}
Re~V_j(x)=a_1a_2,~~Im~V_j(x)=\left(a_1+a_2\right)x_j
-a_1a_2g\sum_{\genfrac{}{}{0pt}{2}{k=1}{k\neq j}}^n
\frac{1}{x_j-x_k},
\end{eqnarray}
and hence the limiting FP operator is
\begin{equation}
L_{FP}=a_1a_2 \left(\sum_{j=1}^n \frac{\partial^2}{\partial x_j^2}
+ 2\sum_{j=1}^n\frac{\partial}{\partial
x_j}\left(\left(\frac{1}{a_1}+\frac{1}{a_2}\right)x_j
-g\sum_{\genfrac{}{}{0pt}{2}{k=1}{k\neq j}}^n
  \frac{1}{x_j-x_k}\right)\right).
\end{equation}
This is just the corresponding FP operator for the $A_{n-1}$ type
CSM system in (\ref{FP-CMS}). The presence of the scale factors
arises from the way the Hamiltonian and the potential of the RSvD
systems are parameterized.  If we take $a_1=a_2=2$, and rescale
 $H\to H/4$, then the two formulas are the same.

\section{Multi-variable $q$-shift Fokker-Planck equation}

As mentioned before, the trigonometric cases can be treated more
elegantly in terms of multiplicative shift type of Hamiltonian
\cite{OS2,OS3}. Let us define $z_j=\exp(2\pi
ix_j/L),~q=\exp(-2\pi/L)$ and $a_0=q^{g_0}$.  For any function
$f(z)$ ($z={}^t(z_1,\cdots,z_n)$) with real coefficients, we have
$f(z)^*=f(z^{-1})$. Furthermore, we define $D_j\equiv
z_j\frac{\partial}{\partial z_j}$. Then $q^{D_j}$ is a
multiplicative $q$-shift operator, i.e.
$q^{D_j}f(z)=f(z_1,\cdots,qz_j,\cdots,z_n)$.  With these, the
Hamiltonian (\ref{H_RS}) of the trigonometric case with
(\ref{V_RS}), (\ref{trigBC_RS_v}) and (\ref{trigBC_RS_w}) can be
recast into (up to an overall constant factor)
\begin{eqnarray}
  H =\sum_{j=1}^n \left(\sqrt{V_j(z)}\,q^{D_j}\!\sqrt{V_j(z^{-1})}
     +\sqrt{V_j(z^{-1})}\,q^{-D_j}\!\sqrt{V_j(z)}
 -V_j(z) - V_j(z^{-1})\right),
  \label{H-q}
\end{eqnarray}
where the potential $V_j(z)$ have the form
\begin{subequations}
\begin{align}
 \text{(\romannumeral3)}\,:\,\ &
 V_j(z)=\prod_{k\neq j}
\frac{1-a_0z_j/z_k}{1-z_j/z_k},
\label{A_Vj}\\
 \text{(\romannumeral4)}\,:\,\ &
 V_j(z)=\frac{\prod_{\alpha=1}^4 (1-a_\alpha
z_j)}{(1-z_j^2)(1-qz_j^2)}\cdot \prod_{k\neq j}\prod_{\epsilon=\pm
1} \frac{1-a_0z_jz_k^\epsilon}{1-z_jz_k^\epsilon}, \label{Vj}
\end{align}
\end{subequations}
and the $\{a_\alpha\}$'s are related to the parameters in
(\ref{trigBC_RS_v}) and (\ref{trigBC_RS_w}) (see \cite{OS1}):
\[
(a_1,a_2,a_3,a_4)=(e^{-{2\pi{g}_1}/{L}},e^{-{2\pi({g}_2+1/2)}/{L}},
   -e^{-{2\pi{g}'_1}/{L}},-e^{-{2\pi({g}'_2+1/2)}/{L}}).
\]
The form of the Hamiltonian (\ref{H-q}) is so chosen that it looks
similar to that of the rational case (\ref{H_RS}) by the formal
replacement $e^{-i\partial_j}\to q^{D_j}$, $e^{i\partial_j}\to
q^{-D_j}$. However, it should be emphasized that the potential
$V_j(z)$ in (\ref{H-q}) corresponds to $V_j^*(x)$ in (\ref{H_RS}),
which is due to the fact that $0<q<1$ ($q=e^{-{2\pi}/{L}}$).

  The ground state
$\phi_0$ satisfying $H\phi_0=0$ is
\begin{subequations}
\begin{align}
\text{(\romannumeral3)}\,:\,\ &
  \phi_0(z\,;g_0,q)\propto \biggl|
  \prod_{1\leq j<k\leq n}
  \frac{(z_j/z_k;q)_{\infty}}{(a_0z_j/z_k;q)_{\infty}}\,
  \biggr|,\\
  \text{(\romannumeral4)}\,:\,\ &
    \phi_0(z\,;\{a_\alpha\},g_0,q)\propto \biggl|\, \prod_{j=1}^n
  \frac{(z_j^2;q)_{\infty}}{\prod_{\alpha=1}^4(a_{\alpha}z_j;q)_{\infty}}
  \,\cdot\!\!\!\!
  \prod_{1\leq j<k\leq n}\prod_{\epsilon=\pm 1}
  \frac{(z_jz_k^{\epsilon};q)_{\infty}}{(a_0z_jz_k^{\epsilon};q)_{\infty}}\,
  \biggr|.
\end{align}
\end{subequations}
Here we use the standard notation
$(a;q)_\infty=\prod_{n=0}^\infty(1-a q^n)$.

Just like the rational cases, the Hamiltonian (\ref{H-q}) is a sum
of factorized forms, i.e. $H=\sum_j A_j^{\dagger}A_j$, with
\begin{eqnarray}
 A_j&=&
 q^{\frac{D_j}{2}}\sqrt{V_j(z^{-1})}
  -q^{-\frac{D_j}{2}}\sqrt{V_j(z)},\\
  A_j^{\dagger}&=&
 \sqrt{V_j(z)}\,q^{\frac{D_j}{2}}
  -\sqrt{V_j(z^{-1})}\,q^{-\frac{D_j}{2}}.
\end{eqnarray}
The ground state $\phi_0$ is annihilated by $A_j$: $ A_j\phi_0
(z)=0~ (j=1,\ldots,n)$.  Explicitly this equation reads
\begin{gather}
  \sqrt{V_j(z_1^{-1},\cdots, q^{-\frac12}z_j^{-1},\cdots,z_n^{-1})}\,
  \phi_0(z_1,\cdots, q^{\frac12}z_j,\cdots,z_n)\nonumber\\
  =\sqrt{V_j(z_1,\cdots, q^{-\frac12}z_j,\cdots,z_n)}\,
  \phi_0(z_1,\cdots, q^{-\frac12}z_j,\cdots,z_n).
  \label{Cond-q}
\end{gather}
Using $H$, $\phi_0$,  and (\ref{Cond-q}), the similarity
transformation (\ref{FP}) produces the FP operator
\begin{eqnarray}
  L_{FP}=- \sum_{j=1}^n \left(q^{-D_j}V_j(z) +
  q^{D_j}V_j(z^{-1}) - V_j(z) - V_j(z^{-1})\right).
  \label{L-q}
\end{eqnarray}
This is the general discrete $q$-deformed FP operator
corresponding to the Hamiltonian (\ref{H-q}).  Again, $L_{FP}$
annihilates $\phi_0^2$.

\section{Summary}

In this paper we have proposed new types of multi-variable FP
equations associated with the RSvD systems which are difference
instead of differential equations.  As the RSvD systems are the
integrable ``discrete" counterparts of the celebrated exactly
solvable many-body CMS systems, the deformed FP equations
presented here  can be considered as ``discrete" deformations of
the ordinary multi-variable FP equations. They are also
integrable.  In the small momentum  (large $c$) limit, our FP
equations reduce to the usual FP equations associated with the
respective CMS systems.  For future study, it would be of interest
to investigate solutions of these generalized FP equations with
potentials $V(x)$ and $V(z)$ other than those exactly solvable
ones discussed here.

\begin{acknowledgments}

This work is supported in part by the National Science Council of
the Republic of China under Grant NSC 95-2112-M-032-012 (CLH), and
in part by the Grant-in-Aid for Scientific Research from the
Ministry of Education, Culture, Sports, Science and Technology of
Japan under Grant No.18340061 and No. 16340040 (RS). We would like
to thank Koryu-kyokai (Japan) and  National Center for Theoretical
Sciences (Taipei) for support through the Japan-Taiwan
collaboration programs.  The authors  thank S. Odake for useful
comments. CLH and RS would like to thank the staff and members of
the Yukawa Institute for Theoretical Physics at the Kyoto
University and the National Taiwan University, respectively, for
their hospitality and financial support during their respective
visits.

\end{acknowledgments}



\begin{thebibliography}{99}

\bibitem{FP}
H. Risken, {\sl The Fokker-Planck Equation} (2nd. ed.)
(Springer-Verlag, Berlin, 1996).

\bibitem{FracFP}
 B. O'Shaughnessy and I. Procaccia, Phys. Rev.
Lett. {\bf 54}, 455 (1985); R. Metzler, E. Barkai and J. Klafter,
Phys. Rev. Lett. {\bf 82},  3563 (1999); Europhys. Lett. {\bf 46},
431 (1999).

\bibitem{NFP}
C. Tsallis and D.\,J. Bukman, Phys. Rev. E {\bf 54},  R2197
(1996); A. Compte and D. Jou, J. Phys. A  {\bf 29}, 4321 (1996).

\bibitem{Ho}
C.-L. Ho, Phys. Lett. A {\bf 268}, 217 (2000).

\bibitem{HS} C.-L. Ho and R. Sasaki, ``Deformed Fokker-Planck
equations", cond-mat/0612318.

\bibitem{OS1}
S. Odake and R. Sasaki, J. Nonlinear Math. Phys. {\bf 12}, S507
(2005); ``Equilibrium positions and eigenfunctions of shape
invariant (`discrete') quantum mechanics", Rokko Lectures in
Mathematics (Kobe University) {\bf 18}, 85 (2005),
  ({\sl Elliptic Integrable Systems}, Eds. M.~Noumi and K.~Takasaki),
hep-th/0505070.

\bibitem{OS2}
S. Odake and R. Sasaki, J. Phys. A {\bf 37}, 11841 (2004); J.
Math. Phys. {\bf 46}, 063513 (2005).

\bibitem{OS3}
S. Odake and R. Sasaki, Prog. Theor. Phys. {\bf 114}, 1245 (2005).

\bibitem{AAR}
G.\,E. Andrews, R. Askey and R. Roy, {\sl Special Functions}
(Cambridge Univ. Press, Cambridge, 1999).

\bibitem{KS}
R. Koekoek and R.\,F. Swarttouw, ``The Askey-scheme of
hypergeometric orthogonal polynomials and its $q$-analogue",
math.CA/9602214.

\bibitem{AW}
R. Askey and J. Wilson, Memoirs Amer. Math. Soc. {\bf 319},
(1985).

\bibitem{RSD}
S.\,N.\,M. Ruijsenaars and H. Schneider, Ann. Phys. {\bf 170}, 370
(1986); S.\,N.\,M. Ruijsenaars, Comm. Math. Phys.  {\bf 110}, 191
(1987); J.\,F. van Diejen, J. Math. Phys. {\bf 35}, 2983 (1994);
J. Phys. A {\bf 28}, L369 (1995); J. Math. Phys. {\bf 36}, 1299
(1995).

\bibitem{CSM}
F. Calogero, J. Math. Phys. {\bf 12}, 419 (1971); B. Sutherland,
Phys. Rev.  A {\bf 5}, 1372 (1972); J. Moser, Adv. Math. {\bf 16}
(1975), 197; Lecture Notes in Phys. {\bf 38} (Springer-Verlag,
1975).

\bibitem{zero}
F. Calogero, Lett. Nuovo Cim. {\bf 22}, 251 (1977), E. Corrigan
and R. Sasaki, J. Phys. A {\bf 35}, 7017 (2002).

\bibitem{Dyson}
F.\,J. Dyson, J. Math. Phys. {\bf 3}, 1191 (1962); ibid. {\bf 13},
90 (1972); M.\,L. Mehta, {\sl Random Matrices}, 3rd Ed. (Elsevier,
San Diego, 2004).

\bibitem{n-s}
O. Narayan and B.\,S. Shastry, Phys. Rev. Lett. {\bf 71}, 2106
(1993); A. Pimpinelli, Phys. Rev. Lett. {\bf 95}, 246101 (2005).
\end{thebibliography}
\end{document}